\documentclass[a4paper]{article}

\usepackage{INTERSPEECH2022}
\usepackage{multirow}
\usepackage[ruled,vlined]{algorithm2e}
\newcommand\tab[1][12pt]{\hspace*{#1}}

\title{Small Footprint Multi-channel ConvMixer for Keyword Spotting with Centroid Based Awareness}

\name{Dianwen Ng$^{1,2}$, Jin Hui Pang$^2$, Yang Xiao$^2$, Biao Tian$^1$, Qiang Fu$^1$, Eng Siong Chng$^2$}
\address{
  $^1$Alibaba Group, Beijing, China\\
  $^2$Nanyang Technological University, Singapore, Singapore}
\email{
\{dianwen.ng, tianbiao.tb, fq153277\}@alibaba-inc.com \\
\{pang0208, yxiao009\}@e.ntu.edu.sg \\
aseschng@ntu.edu.sg}

\begin{document}
\maketitle
\begin{abstract}
It is critical for a keyword spotting model to have a small footprint as it typically runs on-device with low computational resources. However, maintaining the previous SOTA performance with reduced model size is challenging. In addition, a far-field and noisy environment with multiple signals interference aggravates the problem causing the accuracy to degrade significantly. In this paper, we present a multi-channel ConvMixer for speech command recognitions. The novel architecture introduces an additional audio channel mixing for channel audio interaction in a multi-channel audio setting to achieve better noise-robust features with more efficient computation. Besides, we proposed a centroid based awareness component to enhance the system by equipping it with additional spatial geometry information in the latent feature projection space. We evaluate our model using the new MISP challenge 2021 dataset. Our model achieves significant improvement against the official baseline with a 55\% gain in the competition score (0.152) on raw microphone array input and a 63\% (0.126) boost upon front-end speech enhancement. 
\end{abstract}

\section{Introduction}

Voice assistant application in smart devices is getting more widely adopted with the recent success of automatic speech recognition. Keywords such as ``Alexa" or ``Hey Siri" are some of the commonly chosen voice commands used in activating such hands-free application. Likewise, the process of detecting these predetermined words in a continuous utterance is known as keyword spotting (KWS). Primarily, low latency is the key to building a good KWS system as it runs typically on-device. Recent works on small footprint KWS \cite{majumdar2020matchboxnet, rybakov2020streaming, de2018neural} have shown to perform notably well under clean and close-talking audio sets. However, it is observed to deteriorate significantly on the far-field utterance. The decline in performance is prominent in the multi-talker environment with a low signal to noise ratio (SNR). Conventional techniques, including multi-conditioning \cite{sainath2015convolutional, wang2017trainable} and front-end enhancement \cite{weninger2015speech, kolbaek2017multitalker, chen2017deep}, are used to mitigate this observed phenomenon with the former to condition the KWS model with the background noises and the latter to filter the signals of interference from the noisy stream before passing it to the KWS system. Despite these prior methods, multi-conditioning is poor in adapting a broader range of noises \cite{CL2}. Furthermore, the reverberation and multiple sources of interference in far-field speech processing blur the spectral cues that adversely affects the quality of the single-channel speech enhancement.

Multi-channel systems have been studied extensively for improving the noise robustness of speech recognition. Such a system is predominantly deployed for speech enhancement with algorithms like beamforming, noise suppression and localization to improve the enhancement element of the previous single-channel model \cite{li2016neural, yong2017real}. The latest advances \cite{sainath2017multichannel, ochiai2017unified} incorporate neural computation that resembles beamforming in the deep neural networks allowing for joint optimization of multi-channel enhancement and acoustic modelling. In addition, neural-based denoising \cite{tawara2019multi, chen2018building, wang2014training, xu2014regression, hershey2016deep} takes the raw microphone array and learns the multi-channel filtering and feature representation through supervised training. However, small footprint models often face incompetency in learning proficient spatial filtering and noise-robust feature extraction from the raw microphone data. Hence, it remains a challenge for an on-device small-footprint keyword spotting model to function smoothly under a noisy and far-field environment. In response, \cite{wu2020small} uses a three-dimensional single value decomposition filter layer in their low latency model architecture to handle raw microphone array for on-device multi-channel KWS. The proposed structure has displayed substantial improvement over individual single-channel models. Though, the computation for the decomposition is fairly complex. Besides, \cite{ochiai2017unified, 9053989} have established the effectiveness of channel attention in a unified networks for speech enhancement. Nevertheless, it is not frequently explored in small footprint KWS due to the larger computational memory. 

In this paper, we extend the previous work in \cite{ng2022convmixer} to build a novel small footprint multi-channel ConvMixer for keyword spotting. We are motivated by the achievement in noise robustness with attention \cite{ochiai2017unified}, and carefully consider the constraint of computational latency to propose the use of convolution-mixer module in place of the attention unit. The proposed module offers a strong alternative to the attention networks as suggested in \cite{ng2022convmixer} by computing the weighted feature interaction of the global channel to allow the flow of information with varying importance. This allows our model to receive the audio features from other frames and inject this knowledge to prioritize the relevant attributes for the KWS task while achieving more superior features over the noisy and far-field conditions. Most importantly, it is highly efficient with lower memory and computational usage. In addition, we suggest to provide the awareness of the latent space geometry so as to introduce the spatial inductive bias of the networks to boost the prediction performance. Lastly, we present a supplementary investigation at the last section of the experiments to further explore the potential of our proposed model by deploying a frontend beamformer in a multi-look setting. Instead of utilizing a raw microphone array, we recognize the benefits of using a frontend enhancement to improve the noise robustness in our KWS modelling \cite{ji2020integration, yu2020end}. For simplicity, we adopt the conventional mask-based minimum variance distortionless response (MVDR) beamformer to minimize the distortion. In addition, weighted prediction error (WPE) \cite{yoshioka2012generalization} is applied prior to beamforming to dereverberate the acoustic signals based on long-term linear prediction. We show in our experimental results that our model can overcome the high level of residual noise derived from the beamforming to achieve a substantial improvement in our task.

\section{Methodology}

\begin{figure*}[t]
  \centering
  \includegraphics[width=0.99\linewidth, height=6.2cm]{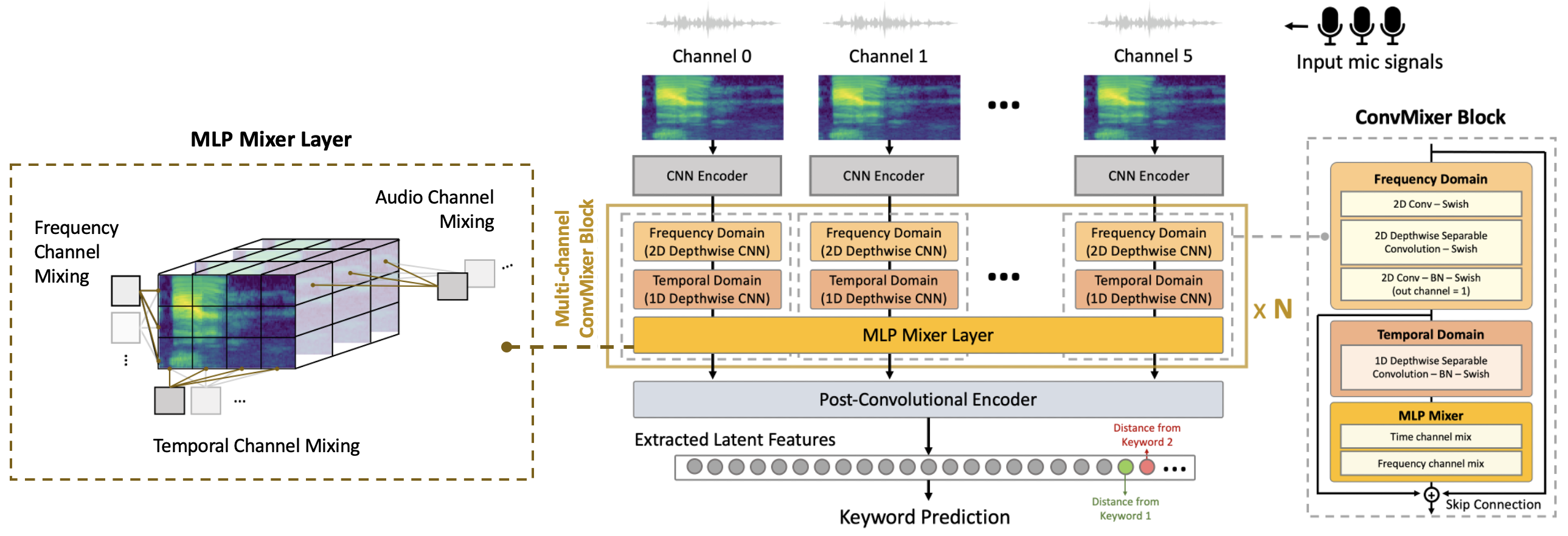}
  \caption{Model Architecture for Multi-channel ConvMixer with Centroid Awareness (An example of a 6-channel model)}
  \label{fig:model}
  \vspace{-0.1cm}
\end{figure*}

\subsection{KWS Model: Multi-channel ConvMixer}
Our proposed model composed of multiple independent single-channel-like ConvMixer. The top portion of the convolutional encoder is similar to \cite{ng2022convmixer}. The middle section is a unique multi-channel convolution-mixer block, and the final part is the post-convolutional block where we extract our latent representation for the class prediction. The small footprint KWS model takes a raw microphone array and converts it into the spectrogram before sending it to the convolutional encoder. Subsequently, they are processed by the multi-channel convolution-mixer block where we implemented a frequency domain, followed by the temporal domain depthwise separable convolution. However, instead of mixing only the tokens--frequency and temporal wise in the mixer layer, a third mixer is introduced that mixes the array of the microphone channel. The following is defined as 
\begin{equation}
\begin{split}
\label{eq:0}
u_{*, t, *} &= x_{*, i, *} + W_2 \cdot \delta[W_1 \cdot \text{LayerNorm}(x)_{*, t, *}] \\
y_{f, *, *} &= u_{f, *, *} + W_4 \cdot \delta[W_3 \cdot \text{LayerNorm}(u)_{f, *, *}] \\
z_{*, *, c} &= y_{*, *, c} + W_6 \cdot \delta[W_5 \cdot \text{LayerNorm}(y)_{*, *, c}]
\end{split}
\end{equation} where $\delta$ represents the GELU unit. $W_1$ and $W_2$ are the learnable weights of the linear layers for temporal channel shared across all frequency $f$, for $f \in \{1, F\}$. $W_3$ and $W_4$ are the learnable weights of the linear layers for frequency channel shared across all $t$, for $t \in \{1, T\}$. And $W_5$ and $W_6$ are the learnable weights of the linear layers for audio channel shared across all $c$, for $c \in \{1, 6\}$ in a six-channel KWS model.

From (\ref{eq:0}), we prompt our model to connect and reference its latent attribute to the features of other channels, i.e. temporal, frequency and microphone, to maximize the effective information for a more informative and robust feature extraction. After that, the process for the multi-channel ConvMixer repeats for a total of $N=4$, carrying the previous information from all other channels to enhance the next individual frequency and temporal domain feature learning of its own track. Finally, the post-convolutional encoder takes the features from all channels and aggregate them with a convolutional layer to obtain a $D$-dimensional vector. To acquire the logits of our class probability map, we append the latent vector with the centroid based awareness as discussed in the next subsection as we send it to a fully connected layer acting as the predictor head.

\subsection{Centroid Based Awareness}
In the standard procedure, the keyword classifier takes the spectrograms of the raw microphone array and learns the embedding function that maps the input to the projecting space. 
Generally, the representation does not hold any additional knowledge about the pre-existing clusters, let alone the geometry of the cluster distribution in the high dimensional space. Suppose if we know the definite representation of our predefined keywords, we believe that inclusion of the awareness for the affinity of the predicting utterance to the keywords will benefit the predictor with the added information about the spatial relationship between the targeted keywords and improve the model fitting especially for the distorted noisy audio. We can do this by measuring the L2-norm Euclidean distance from the input to the keywords, where the keywords are best estimated by the derivation of centroid vector in the cluster.

Formally, we can show this by simply considering the optimization process of the learning loss function. The cross-entropy loss in a general case 
$$
H(q, \hat{q}) = \sum_{k=1}^{C} q^{(k)} \log( \hat{q}^{(k)})
$$ can be decomposed based on Bregman divergence \cite{pfau2013generalized} as

\begin{equation}
\label{eq:1}
\mathbb{E}(H(q, \hat{q})) = \underbrace{D_{\text{KL}}(q || \bar{q})}_{\text{Bias}^2} + \underbrace{\mathbb{E}(D_{\text{KL}}(\bar{q} || \hat{q}))}_\text{Variance}
\end{equation} where $q$ is a one-hot label, $\hat{q}^{(k)}$ is the $k$-th element of the class prediction, and $\bar{q}$ is the average of log-probability after normalization, i.e.

$$
\bar{q}^{(k)} \propto \exp \{\mathbb{E}(log(\hat{q}^{(k)}))\} \text{, for } k = 1,..,C
$$

\noindent Here, the model error consists of the two KL divergence terms in (\ref{eq:1}) and the predictive outcome of $\hat{q}$ is given by the softmax of the dense layer
\begin{equation}
\label{eq:2}
\hat{q} \propto \exp(W_{feat} X_{feat} + W_{L_2\text{-norm}} X_{L_2\text{-norm}})
\end{equation} where $X_{feat}$ is the $D$-dimensional latent representation of $X$, $X_{L_2}$ is the L2-norm of $X_{feat}$ to the keywords. Given that L2-norm measures the similarity distance of the projected input to the keywords and this supplements the computation of the class probability map with the new spatial inductive bias, we should expect $\hat{q}$ to be more confident and stable, where the divergence of $\bar{q}$ and $\hat{q}$ to decrease. Similarly, the average class probability map of $\bar{q}$ gets closer to the label and the divergence of $q$ and $\bar{q}$ decreases. This reduces the estimation bias and variance of our model as presented in (\ref{eq:1}), which converges to a better minimum. Alternatively, this can also be regarded as an inner ensembling (i.e. stacking method) as shown in (\ref{eq:2}) by using the supervised keyword features with L2-norm of the unsupervised K-means clustering.  

To allow for joint optimization, we proposed to compute the centroid to our keywords using the gradient descent algorithm \cite{ruder2016overview}. Specifically, estimating the keyword centroid with the training data is computationally expensive as we would collect the updated latent features of our samples in every training iteration. Therefore, we can utilize stochastic gradient descent by initializing trainable embedding vectors to estimate the centroid. Then, we extract the latent features of every minibatch and update the embedding vectors by minimizing the mean square error (MSE) within the class labels. This is equivalent to deriving the mean of the cluster in the training samples where the sum of error is the lowest. The details are shown in Algorithm \ref{alg:SGD} upon composing the two centroids (i.e. positive and negative) in the binary classification problem.

\begin{algorithm}[ht]
\KwIn{Raw microphone array, $S = \{x_i, y_i\}_{i=1}^N$}
\nl\textbf{Initialize:} \text{Trainable keyword embeddings,} $V_{0,1}$;\\

\nl \textbf{foreach} minibatch, $m = 1, 2, \ldots$ M \textbf{do}\\
\nl \tab $L_{k_0} = \sum_{i \in {k_0}} ||F(x_i) - V_0||^2 $; \hspace{0.03cm} Negative keyword\\
\nl \tab $V_0 \leftarrow V_0 - \eta \nabla L_{k_0}$ \\ \vspace{0.2cm}
\nl \tab $L_{k_1} = \sum_{i \in {k_1}} ||F(x_i) - V_1||^2 $; \hspace{0.03cm} Positive keyword\\
\nl \tab $V_1 \leftarrow V_1 - \eta \nabla L_{k_1}$ 
\caption{\bf Finding Centroid with SGD} 
\label{alg:SGD}
\end{algorithm}
\vspace{-0.01cm}

\section{Experiments}
\subsection{Experimental Setup}
\subsubsection{Keyword Spotting Dataset}

We perform our experiments using the task 1 dataset from the MISP challenge 2021 \cite{chen2022misp}. In our work, we only consider the audio data, where we will build a KWS model that is robust to the
home TV scenario, i.e. noisy and far-field. In particular, a family of people will be seated 3-5m away before the TV, and there may be conversations while someone is interacting with the television. A linear microphone array (6 channels) is placed near the TV at the distance of a far-field (3-5m) condition, and our task aims to detect the following keyword “Xiao T Xiao T” from the recorded utterance. In addition, parallel recordings for mid-field (1-1.5m, 2 channels) and near-field close-talking (1 channel) are provided. The statistical summary of the dataset can be found in Table \ref{tbl:1}.

\begin{table}[ht] \centering\footnotesize
\renewcommand{\arraystretch}{1.2}
\begin{tabular}{|c|cc|cc|c|c|}
\hline
\multirow{2}{*}{Dataset} & \multicolumn{2}{c|}{Train} & \multicolumn{2}{c|}{Dev} & \multirow{2}{*}{Eval} & \multirow{2}{*}{Total} \\ \cline{2-5}
             & \multicolumn{1}{c|}{Pos} & Neg & \multicolumn{1}{c|}{Pos} & Neg &      &        \\ \hline
Duration (h) & \multicolumn{1}{c|}{5.67}     & 112.86   & \multicolumn{1}{c|}{0.62}     & 2.77     & 2.87 & 124.79 \\ \hline
Session      & \multicolumn{2}{c|}{89}                  & \multicolumn{2}{c|}{10}                  & 19   & 118    \\ \hline
Room         & \multicolumn{2}{c|}{25}                  & \multicolumn{2}{c|}{5}                   & 8    & 38     \\ \hline
Participant  & \multicolumn{2}{c|}{258}                 & \multicolumn{2}{c|}{35}                  & 54   & 347    \\ \hline
Male         & \multicolumn{2}{c|}{81}                  & \multicolumn{2}{c|}{11}                  & 31   & 123    \\ \hline
Female       & \multicolumn{2}{c|}{177}                 & \multicolumn{2}{c|}{24}                  & 23   & 224    \\ \hline
\end{tabular}
\caption{Summary of the MISP Challenge 2021, Task 1 Dataset}
\label{tbl:1}
\vspace{-0.6cm}
\end{table}

\subsubsection{Implementation Details}
\textbf{Input Feature} We convert our wav utterance to a 40-dimensional log Mel filterbank (FBank) with a 32ms window size and a 10ms shift. We fixed the length of our FBank at 2s. The shorter utterance will be right-padded with zeros. During training, data augmentation is performed with a random time shift of range between $-100$ to $100$ms. Furthermore, spec augmentation \cite{specaugment} is applied with two--frequency and time maskings of 25 and 7. 

\noindent\textbf{Model Training} Several experiments of single and multi-channel KWS ConvMixer are designed to demonstrate our work. All models are trained on a batch size of 64 and an initial learning rate of 6e-4. The learning rate decays with cosine annealing where the minimum is set to 1e-12. Adam optimizer and binary cross-entropy loss are used in the optimization process. To account for data imbalance, we utilized oversampling strategy during training to augment our positive samples. We trained our model with curriculum learning of three phases, where the first phase uses near-field followed by mid-field and finally the far-field dataset. All training samples contain the original background noise and no additional noise perturbation is added during the augmentation. Moreover, we set our experimental baseline with the architecture of a CNN-LSTM \cite{arik2017convolutional} networks provided by the official. The training configurations for our baseline are unmodified to reproduce the official result.

\subsection{Results}
\begin{table*}[!htbp] \centering\footnotesize
\renewcommand{\arraystretch}{1.1}
\begin{tabular}{|c|c|cccc|cccc|}
\hline
\multirow{2}{*}{Models} &
  \multirow{2}{*}{Params (K)} &
  \multicolumn{4}{c|}{Development} &
  \multicolumn{4}{c|}{Evaluation} \\ \cline{3-10} 
 &
   &
  \multicolumn{1}{c|}{FAR} &
  \multicolumn{1}{c|}{FRR} &
  \multicolumn{1}{c|}{Score} &
  Acc (\%) &
  \multicolumn{1}{c|}{FAR} &
  \multicolumn{1}{c|}{FRR} &
  \multicolumn{1}{c|}{Score} &
  Acc (\%) \\ \hline
Baseline- Official (ch0) \cite{chen2022misp}&
  2,682 &
  \multicolumn{1}{c|}{0.181} &
  \multicolumn{1}{c|}{0.094} &
  \multicolumn{1}{c|}{0.275} &
  87.7 &
  \multicolumn{1}{c|}{0.261} &
  \multicolumn{1}{c|}{0.083} &
  \multicolumn{1}{c|}{0.344} &
  85.7 \\ \hline
ConvMixer (ch0) \cite{ng2022convmixer} &
  124 &
  \multicolumn{1}{c|}{0.032} &
  \multicolumn{1}{c|}{0.144} &
  \multicolumn{1}{c|}{0.176} &
  94.2 &
  \multicolumn{1}{c|}{0.063} &
  \multicolumn{1}{c|}{0.114} &
  \multicolumn{1}{c|}{0.177} &
  92.7 \\ \hline
Beamformer (MVDR) + ConvMixer &
  124 &
  \multicolumn{1}{c|}{0.056} &
  \multicolumn{1}{c|}{0.088} &
  \multicolumn{1}{c|}{0.144} &
  93.7 &
  \multicolumn{1}{c|}{0.048} &
  \multicolumn{1}{c|}{0.121} &
  \multicolumn{1}{c|}{0.169} &
  93.7 \\ \hline
ConvMixer (6-channel model) &
  415 &
  \multicolumn{1}{c|}{0.050} &
  \multicolumn{1}{c|}{0.074} &
  \multicolumn{1}{c|}{\textbf{0.124}} &
  94.5 &
  \multicolumn{1}{c|}{0.043} &
  \multicolumn{1}{c|}{0.118} &
  \multicolumn{1}{c|}{0.161} &
  94.1 \\ \hline
\begin{tabular}[c]{@{}c@{}} [Ours] Centroid Awareness ConvMixer\\ (6-channel model) $\dagger$\end{tabular} &
  622 &
  \multicolumn{1}{c|}{0.034} &
  \multicolumn{1}{c|}{0.091} &
  \multicolumn{1}{c|}{0.125} &
  \textbf{95.3} &
  \multicolumn{1}{c|}{0.044} &
  \multicolumn{1}{c|}{0.107} &
  \multicolumn{1}{c|}{\textbf{0.152}} &
  \textbf{94.3} \\ \hline
\begin{tabular}[c]{@{}c@{}}Centroid Distance Clustering $\ddagger$ \end{tabular} &
  N.A. &
  \multicolumn{1}{c|}{0.026} &
  \multicolumn{1}{c|}{0.106} &
  \multicolumn{1}{c|}{0.132} &
  95.6 &
  \multicolumn{1}{c|}{0.040} &
  \multicolumn{1}{c|}{0.132} &
  \multicolumn{1}{c|}{0.172} &
  94.1 \\ \hline
\end{tabular}
\caption{Performance of our experimental models with Task 1, MISP challenge 2021 development and evaluation set. \\ $\ddagger$: uses the distance computed from the representation in the projected space of our proposed model $\dagger$ to the centroids and performs a shortest distance clustering}
\label{tbl:2}
\vspace{-0.5cm}
\end{table*}

The experiments are conducted to demonstrate the quantitative improvement of our proposed model against the official baseline and the version of a single-channel small size ConvMixer. We evaluate all models based on the accuracy of the predictive score from the official development (dev) and evaluation (eval) set. In addition, to offset the likelihood of an over-optimistic assessment derived from the highly imbalanced class distribution, we will also analyze our work with the sum between false alarm rate (FAR) and false rejection rate (FRR).  This metric measures the sensitivity of the error coming from the binary classes and is given by
$$\text{Score} = FAR + FRR$$ where the FAR and FRR are defined as follows
$$FAR = \frac{FP}{FP + TN} \hspace{0.5cm}  FRR = \frac{FN}{FN + TP}$$

\begin{figure}[!h]
  \centering
  \includegraphics[width=0.46\textwidth]{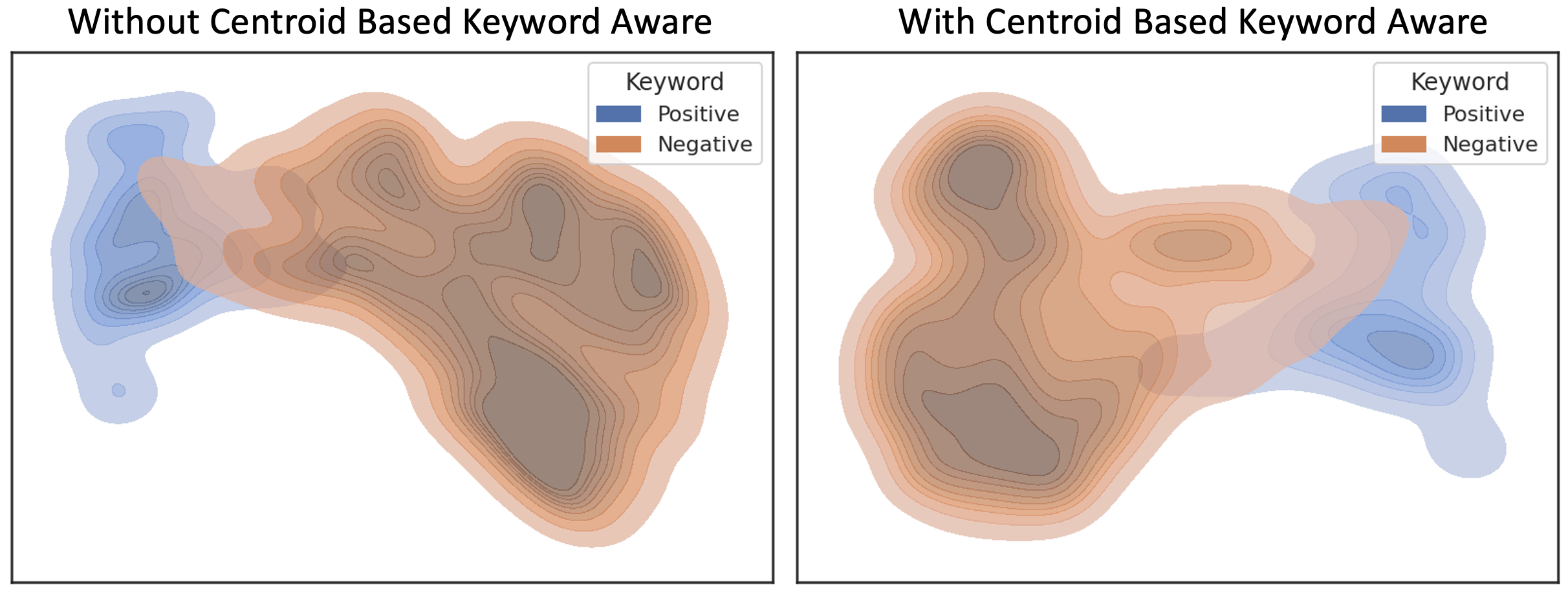}
  \caption{Distributive plot of the evaluation set}
  \label{fig:distplot}
\end{figure}

We first compare the official baseline that resembles a single-channel CNN-LSTM networks with a parameter size of 2.68M. Here, we mainly discuss the performance on eval for a more objective view as the data is held completely independent from the training phase. From Table \ref{tbl:2}, the baseline has achieved a decent performance of 0.34 and 85.7\% accuracy. However, the single-channel ConvMixer with a parameter size of 124K (4.6\% in size of baseline) has evidently outperformed in score with 0.17 (48\%) improvement and 92.7\% (8.2\%) in accuracy. Nonetheless, our proposed multi-channel model (without centroid awareness) has beaten the single-channel ConvMixer with an additional gain of 0.016 (9.0\%) in score and 94.1\% (1.5\%) in accuracy. Our multi-channel model has also proven its superiority, i.e. 4.7\% in score and 0.4\% in accuracy, over the front-end enhanced speech model where we train the single-channel ConvMixer with a naive MVDR beamformer to improve the quality of the input utterance before sending it for prediction. Lastly, as we include the component of the centroid based keyword, the score has an additional boost of 5.6\% and 0.2\% in accuracy. Overall, our proposed final system has achieved a total of 55\% in score and 10\% in accuracy improvement as compared to the official baseline, and an increment of 14\% in score and 1.7\% in accuracy in comparison to the single-channel ConvMixer. Furthermore, we plot the distribution of the latent representations in Figure \ref{fig:distplot} to present a visual understanding of the benefit from adding the centroid aware as we observe lower overlapping of highly concentrated region of the distribution between the binary classes. 

\vspace{-0.5cm}

\subsection{Empirical studies with Multi-look Beamformer and Weighted Predictive Error}
The purpose of this section is to demonstrate the potential of our proposed model in performing the keyword spotting task under far-field and noisy conditions. We discuss some of the commonly used tricks with easy to implement front-end preprocessing to enhance the robustness of our prediction. In particular, we refer to \cite{sainath2017multichannel, ji2020integration, yu2020end} and modify our networks into a multi-look beamforming KWS by simply replacing the raw microphone array with a set of beamformed signals. Each beam signal is designed to have a different look direction. For simplicity, we choose the conventional mask-based minimum variance distortionless response (MVDR) beamformer to minimize the distortion. Furthermore, given the use for a TV scenario, beams are targeted at 10$^\circ$, 90$^\circ$ and 170$^\circ$ respectively. In addition, we include a reference wav signal from the channel 0 raw to preserve the information of the original utterance. As a result, we obtain a multi-look (3-look + ch0) KWS ConvMixer model with centroid awareness. Besides, we also propose to perform dereverberation with WPE and we present our empirical results in Table \ref{tbl:3}.

\begin{table}[ht]\centering\footnotesize
\renewcommand{\arraystretch}{1.2}
\begin{tabular}{|c|ccccc|}
\hline
\multirow{2}{*}{\begin{tabular}[c]{@{}c@{}}Models\\ (Input Audio)\end{tabular}} & \multicolumn{5}{c|}{Evaluation Set}                                                                                             \\ \cline{2-6} 
                        & \multicolumn{1}{c|}{Param} & \multicolumn{1}{c|}{FAR}  & \multicolumn{1}{c|}{FRR}  & \multicolumn{1}{c|}{Score} & Acc \\ \hline
3-look Beamformer       & \multicolumn{1}{c|}{473 K}        & \multicolumn{1}{c|}{0.047} & \multicolumn{1}{c|}{0.090} & \multicolumn{1}{c|}{0.137}  & \textbf{94.5}       \\ \hline

6-channel WPE           & \multicolumn{1}{c|}{622 K}        & \multicolumn{1}{c|}{0.040} & \multicolumn{1}{c|}{0.136} & \multicolumn{1}{c|}{0.176}  & 94.0       \\ \hline

WPE + 3-look Beam & \multicolumn{1}{c|}{473 K}        & \multicolumn{1}{c|}{0.054} & \multicolumn{1}{c|}{0.072} & \multicolumn{1}{c|}{\textbf{0.126}}  & 94.2       \\ \hline
\end{tabular}
\caption{Performance on eval set with front-end processing}
\label{tbl:3}
\vspace{-0.5cm}
\end{table}

From the table, there are additional improvements as we introduced the front-end enhancement into our system. We obtain a score of 0.137 (9.9\%) for our 3-look beamformer with 94.5\% (0.2\%) accuracy in the detection task. However, dereverberation with WPE alone does not help in training a better model. This is likely due to the simplicity of our algorithm that instils adverse distortion when it attempts to blindly dereverberate the signals. Nevertheless, dereverberation seems to help lower our FAR where the model is more careful in assigning positives. On the whole, despite a small fractal cost in the latency for the front-end enhancement, a combination of both preprocessing has substantially improved the score for a total of 63\% in comparison to the baseline. Above all, this is competitive against the competition's leaderboard with our audio-only model of small footprint KWS. At last, we anticipate a further gain with non-linear neural enhancement incorporated within the KWS networks in future work.

\section{Conclusions}

To conclude,  we proposed a novel small-footprint model for multi-channel KWS with ConvMixer module and centroid based awareness. Our model has achieved a compelling gain with a 55\% improvement against the official baseline under a noisy and far-field condition. Additionally, we observed a 63\% boost in the score--0.126 with the added front-end enhancement bringing our model to be competitive against the competition's leaderboard with an audio-only model of parameter size 473K. Besides, this also suggests that our small model has achieved better robustness in noisy and far-field environment.  

\section{Acknowledgements}

This work was supported by Alibaba Group through Alibaba Innovative Research (AIR) Program and Alibaba-NTU Singapore Joint Research Institute (JRI), Nanyang Technological University, Singapore.

\bibliographystyle{IEEEtran}

\bibliography{template}


\end{document}